\documentclass[twocolumn,showpacs,preprintnumbers,amsmath,amssymb]{revtex4}
\usepackage{graphicx}% Include figure files
\usepackage{dcolumn}% Align table columns on decimal point
\usepackage{bm}% bold math

\begin{document}
\title{ Oscillating Nernst-Ettingshausen effect in Bismuth across the quantum limit }
\author{Kamran Behnia$^{1}$, Marie-Aude M\'easson$^{2}$ and Yakov Kopelevich$^{3}$}
\affiliation{(1)Laboratoire de Physique Quantique(CNRS), ESPCI, 10
Rue de Vauquelin,
75231 Paris, France \\
(2)Graduate School of Science, Osaka University, Toyonaka, Osaka,
560-0043 Japan \\
(3)Instituto de Fisica ``Gleb Wataghin'', UNICAMP, 13083-970
Campinas, S\~{a}o Paulo, Brazil}

\date {April 18, 2007}

\begin{abstract}
In elemental bismuth, 10$^5$  atoms share a single itinerant
electron. Therefore, a moderate magnetic field can confine electrons
to the lowest Landau level. We report on the first study of metallic
thermoelectricity in this regime. The main thermoelectric response
is off-diagonal with an oscillating component  several times larger
than the non-oscillating background. When the first Landau level
attains the Fermi Energy, both the Nernst and the Ettingshausen
coefficients sharply peak, and the latter attains a
temperature-independent maximum. These features are yet to be
understood. We note a qualitative agreement with a theory invoking
current-carrying edge excitations.
\end{abstract}


\maketitle

Bismuth crystalizes in a rhombohedral structure, which can be
obtained by distorting a cubic lattice along its body diagonal. As a
consequence of this slight deviation from higher symmetry, it
becomes a semi-metal with a Fermi surface occupying a tiny fraction
of the Brillouin zone\cite{edelman}. It has frequently played a
pioneer role in the history of metal physics. Phenomena such as
giant magnetoresistance, de Hass-van Alphen effect, Shubnikov-de
Hass effect, and oscillatory magnetostriction were all first
observed in bismuth\cite{edelman}. The Nernst effect was also
originally discovered in bismuth\cite{ettingshausen}. The
archetypical semi-metal and its unique properties are attracting new
attention \cite{murakami}, following the recent observation of Dirac
Fermions in graphite\cite{luk,zhou} and
graphene\cite{novoselov,zhang}. Electronic transport in bismuth is
indeed fascinating: Magnetic field can enhance resistivity by a
factor of 10$^6$ without saturation. The quasi-linear
magneto-resistance is incompatible with the semi-classical
theory\cite{abrikosov}. Possible occurrence of exotic field-induced
instabilities ranging from an excitonic insulator\cite{fenton} to a
topological insulator driven by spin-orbit interactions\cite{fu}
continues to fuel research and
debate\cite{bompadre,kopelevich,du,kopelevich2}.

\begin{figure}
\resizebox{!}{0.6\textwidth}{\includegraphics[width=\linewidth]{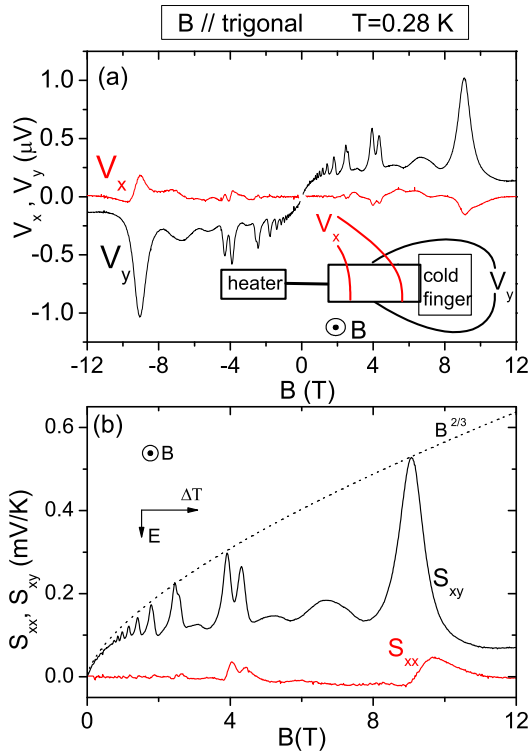}}\caption{a)
Experimental set-up and the longitudinal voltage obtained in
presence of a constant heat current as a function of magnetic field
swept from -12 T to +12 T. b) Nernst and Seebeck coefficients
extracted from the odd [even] component of V$_{y}$[V$_{x}$]. The dot
line representing a $B^{2/3}$ behavior is just a guide for eye.
Arrows representing the three relevant vectors define the sign of
the measured Nernst signal.}
\end{figure}

Previous studies of thermoelectric response of bismuth in quantizing
magnetic fields were performed decades ago and restricted to
temperatures above 4.2 K and to relatively low fields (B $<$5
T)\cite{steele,mangez}. Here, we present a study of
thermoelectricity extending down to  0.28 K and up to 12 T. This
allows to attain the quantum limit, when the magnetic length equals
the Fermi wavelength. Both transverse [Nernst] and longitudinal
[Seebeck] components of the thermoelectric tensor display quantum
oscillations, but the Nernst response dominates by far. The Nernst
signal is large when a Landau level meets the chemical potential and
is damped when the latter lies between two successive Landau levels.
The change in the magnitude is much larger than the quantum
oscillations of the density of states monitored by the Shubnikov-de
Hass effect. These experimental findings are a challenge to our
understanding of weakly-interacting electrons in the quantum limit.
Qualitatively, they are strikingly similar to a recent theoretical
predication invoking edge states in the context of the Quantum Hall
Effect\cite{nakamura}. We also note, but fail to explain, that the
Ettingshausen coefficient becomes temperature-independent in the
low-temperature limit. This offers another notable constraint for
theory.

A longitudinal thermal gradient can produce a longitudinal electric
field (Seebeck effect), a transverse electric field (Nernst effect)
and  a transverse thermal gradient (Righi-Leduc effect). We used a
one-heater-three-thermometers set-up to measure the two
thermoelectric coefficients, the thermal and electric conductivities
as well as the Righi-Leduc coefficient. Because of the dominance of
the lattice thermal conductivity\cite{behnia}, the latter was found
to be very small. Therefore, the measured adiabatic Nernst
coefficient is identical to the isothermal one. The physical
properties of the bismuth single crystal used in this study (RRR=
47, dimensions 0.8$\times$2.2$\times$ 4 mm$^{3}$ ) were presented in
a recent communication by the authors\cite{behnia}. In all
measurements detailed below, the direction of the applied current is
along the binary axis.

The raw data obtained for a field along the trigonal axis at T=0.28
K is shown in Fig. 1. The longitudinal and the transverse voltages
created by applying a constant heat current [which, since the
thermal conductivity is field-independent implies a constant
$\nabla_{x}$T] are plotted as a function of magnetic field swept
from -12 T to +12 T. The transverse signal is by far larger
(reflecting the dominance of the Nernst over the Seebeck response)
and contaminates the longitudinal voltage as evidenced by the sign
reversal of the latter when the field is reversed. Panel b of the
same figure shows the Nernst (S$_{xy}$=E$_{y}$/$\nabla_{x}$T) and
Seebeck(S$_{xx}$=E$_{x}$/$\nabla_{x}$T) coefficients extracted from
the odd [even] component of V$_{y}$(B) [V$_{x}(B)$]. Both show
oscillations, but the transverse component is clearly dominant.

\begin{figure}
\resizebox{!}{0.6\textwidth}{\includegraphics[width=\linewidth]{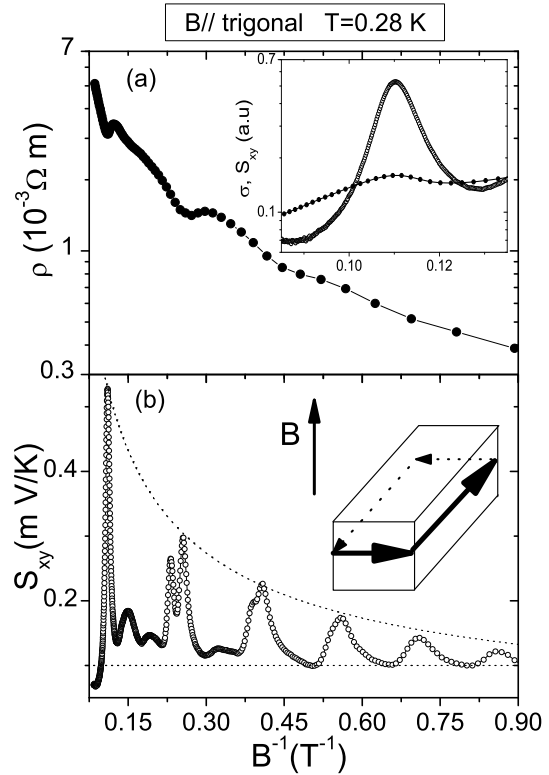}}\caption{a)
The resistivity of the sample as a function the inverse of the
magnetic field.  b) The Nernst signal as a function of the inverse
of the magnetic field. Dotted lines (constant and following
B$^{2/3}$) are guide for eyes. The upper inset compares the changes
in the electric conductivity and the Nernst signal in the vicinity
of the first peak in a semi-log scale. The lower inset represents
wrapping currents in a bulk sample.}
\end{figure}

The sheer magnitude of the oscillations in S$_{xy}$ compared to the
non-oscillating background is remarkable. In bismuth, Shubnikov-de
Hass oscillations of resistivity are easily
observable\cite{bompadre,yang}. However, even at 20 mK, only a
fraction of $\rho_{xx}(B)$ is oscillating and the total resistivity
is dominated by a large non-saturating
magneto-resistance\cite{bompadre}.  Fig. 2 displays both resistivity
and the Nernst response of our sample as a function of the inverse
of the magnetic field, B$^{-1}$. In the case of resistivity, the
passage of successive Landau levels is marked by dips in a
monotonous background (the scale is semi-logarithmic). In the case
of the Nernst response, on the other hand, the oscillations dominate
a flat background. In other words, the anomalies reveal two distinct
regimes : when the chemical potential is located between two
successive Landau levels, S$_{xy}$ is small and is barely
field-dependent. When a Landau level is precisely at the chemical
potential, S$_{xy}$ peaks to a value which is an increasing function
of the magnetic field. The peaks become sharper as one moves towards
the lower Landau levels. The inset of Fig. 2 compares S$_{xy}$ and
the electric conductivity ($\sigma \simeq 1/\rho_{xx}$ since
$\rho_{xy} \ll \rho_{xx}$) in the vicinity of the first peak in a
semi-log plot. Both peak at the same field. However, while $\sigma$
is enhanced by 20 percent, the increase in S$_{xy}$ is multi-fold.
Therefore, the Nernst peak is \emph{not} simply reflecting the
behavior of a transport coefficient when the effective density of
states oscillates in quantizing magnetic fields.

\begin{figure}
\resizebox{!}{0.6\textwidth}{\includegraphics{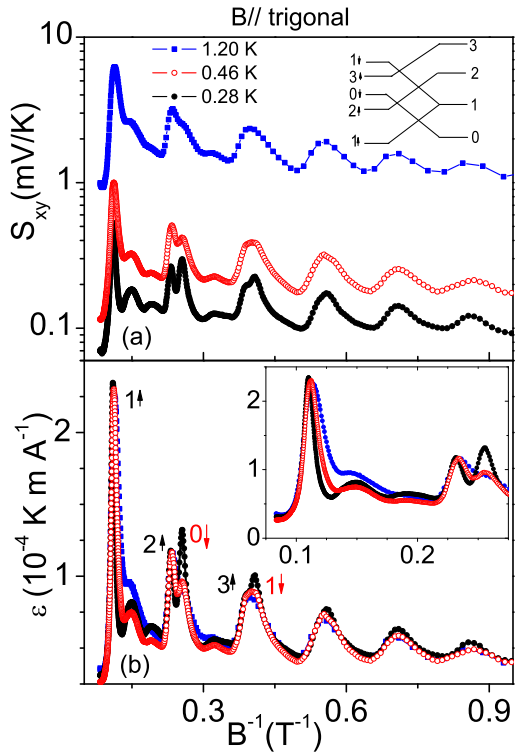}}\caption{(color
online) a: Nernst effect as a function of inverse magnetic field
along the trigonal for three different temperatures. The inset shows
the hierarchy of energy levels. b: The Ettingshausen coefficient
obtained using the same data (See text). The quantum numbers of
first peaks are given. The inset is a zoom on the high-field
regime.}
\end{figure}

Figure 3 presents the Nernst signal at three different temperatures
in a semi-log. plot. The decay in temperature is faster than linear
and roughly T$^2$.  We presented and analyzed the temperature
dependence of the thermal conductivity, $\kappa$, of the same sample
in ref.\cite{behnia} and concluded that heat is almost entirely
transported by ballistic phonons and $\kappa \propto T^{3}$ from the
lowest temperature up to T$\sim$3.8 K. Thanks to the absolute
dominance of the lattice contribution, $\kappa$ remains insensitive
to the magnetic field. The field dependence of the Ettingshausen
coefficient at these three temperatures can be obtained using our
data and the Bridgman relation\cite{bridgman}, which links
Ettingshausen ($\epsilon$) and Nernst(S$_{xy}$) coefficients via
thermal conductivity ($\kappa$). Namely:
\begin{equation}\label{1}
    S_{xy}=\frac{\kappa}{T}\epsilon
\end{equation}

This relation allows to compute $\epsilon$ using  S$_{xy}$ and
$\kappa$. The results are displayed in the main panel of Fig. 3. The
Ettingshausen coefficient is virtually temperature-independent
between 0.28 K and 1.2 K. More precisely, the temperature dependence
of some the peaks saturates below 1.2K. As discussed below, these
peaks are identified as those associated with the Landau level of
holes with spins parallel to field. Explaining their magnitude
appears as a well-defined challenge to the theory.

In contrast to metals, the thermoelectric response of semiconductors
has been extensively studied beyond the quantum limit. In
particular, the thermoelectricity of a two-dimensional electron
gas(2DEG) in a quantizing magnetic field was studied both
theoretically\cite{jonson,oji,cooper} and experimentally
\cite{fletcher,bayot}. Early theoretical studies\cite{jonson,oji}
highlighted the role played by the edge states\cite{halperin} and
predicted sharp features in the thermoelectric response when a
Landau Level attains the Fermi energy. However, in contrast to what
is observed here, the main thermoelectric response was expected to
be diagonal ($S_{xx}$). Only in presence of disorder, the
off-diagonal component (S$_{xy}$) would display a feature
\cite{jonson,oji,cooper}. Experimentally, both S$_{xx}$
\cite{fletcher,bayot} and S$_{xy}$ \cite{fletcher} were found to
show quantum oscillations of comparable magnitude.  More recently,
Nakamura \emph{et al.} presented a new theoretical study of a
ballistic 2DEG in the quantum Hall regime\cite{nakamura}. They
concluded that a circulating edge current would realize a
non-equilibrium steady state and the principal thermoelectric
response would be off-diagonal. In their picture, when the chemical
potential lies between two Landau levels, the edge currents of
ballistic electrons from the hot bath to the cold one (and vice
versa) suppress the Nernst voltage. But when the chemical potential
is precisely located at a Landau level, a finite Nernst response is
restored. Qualitatively, the features observed in our study are
compatible with these predictions.
\begin{figure}
\resizebox{!}{0.52\textwidth}{\includegraphics[width=\linewidth]{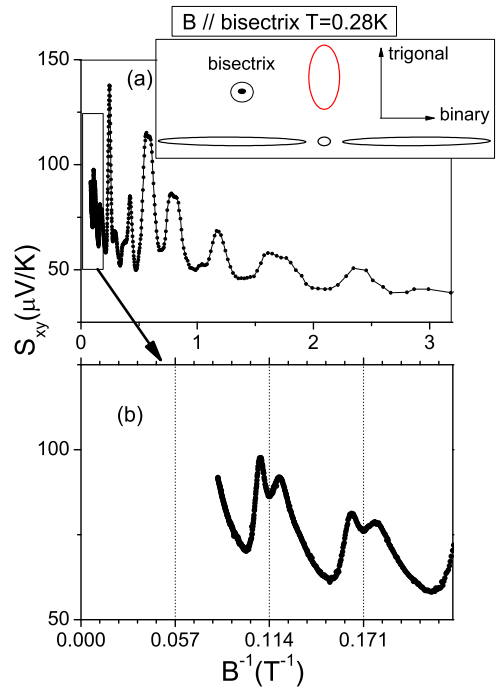}}\caption{a:
Nernst effect as a function of inverse magnetic field along the
bisectrix. The inset shows the cross section of the three electron
ellipsoid and the hole ellipsoid (in red) for this geometry. The
lower panel is a zoom on high-field data which isolates peaks
corresponding to the hole ellipsoid.}
\end{figure}

There is ample evidence for current-carrying edge states in two
dimensions\cite{haug}. In the case of a three dimensional crystal,
Quantum Hall Effect has been observed in a number of bulk
systems\cite{cooper2,elefant,hill} and the associated ``wrapping
currents''\cite{koshino} are experimentally plausible. Note,
however, that in these cases, the metallic surface states emerge in
presence of (and in distinction to) a bulk insulator. In spite of
its exceptionally large magneto-resistance, bismuth is believed to
remain a metal at high magnetic fields and low
temperature\cite{abrikosov,du,kopelevich2}. Moreover, there has been
no report on quantized Hall effect in this three-dimensional system.
Careful studies, designed to distinguish between the bulk and
surface states, are required to shed light on this issue.

An open question is the contribution of phonon drag to the Nernst
signal. A scenario invoking phonons cannot be ruled out. Decades
ago, Gurevich \emph{et al.}\cite{gurevich} argued that the
absorption of acoustical phonons by electrons should sharply peak
when a Landau Level is located at the chemical potential. This is
experimentally confirmed and sound attenuation coefficient of
bismuth display giant quantum oscillations\cite{kajimura}.

While the origin and the magnitude of the giant oscillations
observed here are yet to be understood, their period can be compared
to the well-known topology of the Fermi surface in
bismuth\cite{edelman}. It consists of a hole ellipsoid and three
electron ellipsoids. The volume of the former is equal to the total
volume of the latter. The inset of Fig.4 shows a projection in the
plane normal to the bisectrix (See ref.\cite{yang} for a 3-D
representation). The tiny dimensions of these ellipsoids set
frequencies of the order of a few teslas for quantum oscillations.
The period of the oscillations observed here for a field along the
trigonal (0.147 T$^{-1}$) corresponds to the hole ellipsoid. It is
very close to what was reported by Bompadre \emph{et
al.}\cite{bompadre} (0.146 T$^{-1}$). Moreover, it yields an
equatorial radius of 0.144 nm$^{-1}$ very close to what is expected
($0.146 nm^{-1}$) by band calculations\cite{liu}. As seen in fig. 3,
at high $B/T$ values, the peaks show an additional structure. The
structure can be understood thanks to a careful study by Bompadre
\emph{et al.}\cite{bompadre}, who concluded that the Zeeman energy
is more than twice the cyclotron energy
($E_{Z}=2.16\hbar\omega_{c}$\cite{edelman,bompadre}. This puts the
$|n,\downarrow\rangle$ energy level above $|n+2,\uparrow\rangle$
(See the inset of Fig.3). Remarkably, the Ettingshausen peaks
associated with the minority spins continue to evolve even at 0.28
K.

For a field along trigonal, the electron ellipsoids were not
detectable in previous studies of quantum
oscillations\cite{bhargava,bompadre}. The lower mobility of
electrons compared to holes is attested by their threefold higher
Dingle temperature\cite{bhargava}. Moreover, in this configuration,
the cross section of electron and hole ellipsoids become
accidentally resonant, making the detection of the electron
frequency very difficult.

We also studied the Nernst effect for a field oriented in the plane
normal to trigonal and almost along the bisectrix (the misalignment
was less than 10 degrees). The data are presented in Fig. 4. For
this geometry, the cross section of the hole ellipsoid is larger
than the electron ellipsoids and long-period oscillations of the
latter dominate the Nernst response at lower fields below 6 T. Above
this field, electrons have been all pushed to their lowest Landau
levels and the residual oscillations are due to holes. As seen in
the lower panel of Fig. 4, four peaks corresponding to two
successive Landau levels and their Zeeman splitting can be clearly
identified.  For B $<$6 T, the oscillations due to the three
electron ellipsoids dominate the Nernst response. The longest
detected  period ($\sim$ 0.8 T$^{-1}$) corresponds to a cross
section with an average radius of 0.06 nm$^{-1}$, comparable to the
reported values for the two shorter radii of the cigar-like
electronic ellipsoid (0.05-0.07 nm$^{-1}$\cite{edelman,liu}).

Scrutinizing the \emph{phase} of the quantum oscillations would be
instructive for probing the Berry phase of electrons. According to
Mikitik and Sharlai\cite{mitikik}, when the electrons orbit around a
line of band-contact the semi-classical quantization condition would
be modified. Extending the study to fields larger than 12 T and well
in to the ultra-quantum limit could help to address this point in
bismuth.

We are grateful to H. Aubin, N. Hatano, I. Luk'yanchauk, G. Mikitik
and Y. Sharlai for useful discussions and to Y. Nakajima,  M.
Nardone and A. Pourret for technical assistance. This work was
supported in France by the ICENET project (ANR) and by ECOM-COST P16
(EU) and in Brazil by CNPq and FAPESP.


\begin{thebibliography}{}
\bibitem{edelman} V. S. Edelman, Adv. Phys. \textbf{25}, 555 (1976) and references therein.
\bibitem{ettingshausen} A. V. Ettingshausen and W. Nernst,  Ann. phys. Chem., \textbf{265}, 343 (1886)
\bibitem{murakami} S. Murakami, Phys. Rev. Lett. \textbf{97}, 236805 (2006)
\bibitem{luk} I. A. Luk'yanchuk and Y. Kopelevich, Phys. Rev. Lett. \textbf{93}, 166402 (2004)
\bibitem{zhou} S. Y. Zhou \emph {et al.}, Nature Phys. \textbf{2}, 595 (2006)
\bibitem{novoselov} K. S. Novoselov \emph{et al.}, Nature \textbf{438}, 197 (2005)
\bibitem{zhang} Y. B. Zhang \emph {et al.}, Nature \textbf{438}, 201(2005)
\bibitem{abrikosov}A. A. Abrikosov, J. Phys. A \textbf{36}, 9119 (2003)
\bibitem{fenton} E. W. Fenton, Phys. Rev. \textbf{170}, 816 (1968)
\bibitem{fu} L. Fu, C. L. Kane and E. J. Mele, \textbf{98}, 106803 (2007)
\bibitem{bompadre}S. G. Bompadre \emph{et al.}, Phys. Rev. B \textbf{64}, 073103 (2001)
\bibitem{kopelevich} Y. Kopelevich \emph{et al.}, Phys. Rev. Lett. \textbf{90}, 156402 (2003)
\bibitem{du} X. Du \emph{et al.}, Phys. Rev. Lett. \textbf{94}, 166601 (2005)
\bibitem{kopelevich2} Y. Kopelevich \emph{et al.}, Phys. Rev. B \textbf{73}, 165128 (2006)
\bibitem{steele}M. C. Steele, and J. Babiskin, Phys. Rev. \textbf{98}, 359 (1955)
\bibitem{mangez} J. H. Mangez, J. P. Issi and J. Heremans, Phys. Rev. B \textbf{14}, 4381 (1976)
\bibitem{nakamura} H. Nakamura, N. Hatano and R. Shirasaki, Solid
State Comm. \textbf{135}, 510 (2005)
\bibitem{halperin}B. I. Halperin, Phys. Rev. B \textbf{25}, 2185 (1982)
\bibitem{behnia} K. Behnia, M. -A. M\'easson and Y. Kopelevich, Phys. Rev. Lett. \textbf{98},
076603 (2007)
\bibitem{yang}F. Y. Yang \emph{et al.}, Phys. Rev. B \textbf{61}, 6631 (2000)
\bibitem{bridgman} P. W. Bridgman, Phys. Rev. \textbf{24}, 644
(1924); A. Sommerfeld and N. H. Frank, Rev. Mod. Phys. \textbf{3}, 1
(1931).
\bibitem{fletcher} R. Fletcher, Semicond. Sci. Technol. \textbf{14}, R1 (1999)
\bibitem{jonson} M. Jonson and S. M. Girvin, Phys. Rev. B \textbf{29}, 1939 (1984)
\bibitem{oji} H. Oji, J. Phys. C \textbf{17}, 3059 (1984)
\bibitem{cooper} N. R. Cooper, B. I. Halperin and I. M. Ruzin, Phys. Rev. B \textbf{55}, 2344 (1997)
\bibitem{bayot} V. Bayot \emph{et al.}, Phys. Rev. B \textbf{52}, R8621 (1995)
\bibitem{haug} R. J. Haug, Semicond. Sci. Technol. \textbf{8}, 131 (1993)
\bibitem{cooper2} J. R. Cooper \emph{et al.}, Phys. Rev. Lett. \textbf{63}, 1984
(1989);  S. T. Hannahs \emph{et al.}, Phys. Rev. Lett. \textbf{63},
1988 (1989)
\bibitem{elefant} D. Elefant, G. Reiss and Ch. Baier, Eur. Phys. J. B \textbf{4} , 45 (1998)
\bibitem{hill}S. Hill \emph{et al.}, Phys. Rev. B \textbf{58}, 10778 (1998)
\bibitem{koshino} M. Koshino, H. Aoki and B. I. Halperin, Phys. Rev. B \textbf{66}, 081301(R) (2002)
\bibitem{gurevich}V. L. Gurevich \emph{et al.}, JETP Lett. \textbf{13}, 552(1961)
\bibitem{kajimura} K. Kajimura \emph{et al.}, Phys. Rev. B \textbf{12} (1975)
\bibitem{liu} Yi Liu and R. E. Allen, Phys. Rev. B \textbf{52}, 1566 (1995)
\bibitem{bhargava} R. N. Bhargava, Phys. Rev. \textbf{156}, 785 (1967)
\bibitem{mitikik} G. P. Mikitik and Yu. V. Sharlai, Phys. Rev. Lett. \textbf{82}, 2147(1999)
\end{thebibliography}
\end{document}